\documentstyle[11pt,psfig]{article}
\newcommand{\s}{\rm}

\newcommand{\ra}{\rightarrow}
\newcommand{\mn}{\mu \nu}

\newcommand{\be}{\begin{equation}}
\newcommand{\ee}{\end{equation}}

\newcommand{\bea}{\begin{eqnarray}}
\newcommand{\eea}{\end{eqnarray}}
\newcommand{\bef}{\begin{figure}}
\newcommand{\eef}{\end{figure}}

\newcommand{\rpp}{\rho \pi \pi}

\newcommand{\lgl}{\langle}
\newcommand{\rgl}{\rangle}

\begin{document}
\begin{center}
\large{\bf Low Mass Dileptons from Pb+Au Collisions at CERN SPS}
\vskip 0.2 in
Sourav Sarkar$^1$, Jan-e Alam$^2$ and T. Hatsuda$^2$
\vskip 0.1 in

\small{\it $^1$~Variable Energy Cyclotron Centre,
     1/AF Bidhan Nagar, Calcutta 700 064
     India}

\small{\it $^2$~Physics Department, University of Tokyo, Tokyo 113-0033, Japan}
\end{center}

\begin{abstract}
We show that the dilepton spectra measured by the
CERES collaboration in Pb + Au interactions 
for various charge multiplicities can be 
reproduced by a hadronic initial state 
with reduction in the masses 
of the vector mesons 
in the thermal bath.
Though such an effect may also be achieved 
by a large broadening of the spectral function
we show that the photon spectra is insensitive to this.
It is found that for higher multiplicities  
a good description of the 
data can also be obtained with quark gluon plasma 
initial state if the reduction of the vector
meson masses in the  mixed and hadronic phases
is taken into account. 
We observe that a thermal source with initial temperature
$\sim 200$ MeV can reproduce the observed
enhancement in the low mass region of the dilepton spectra. 
It is not possible to state
which one of the two initial states (QGP or hadronic) 
is compatible with the data.
These findings are in agreement with our
earlier results obtained 
from the analysis of the WA98 photon spectra.
We estimate the number of $\pi-\pi$
collisions near the $\rho$-peak  
of the dilepton spectra and argue that  
thermal equilibrium may have been achieved 
in the system, justifying the use of hydrodynamical
model to describe the space time evolution.

\end{abstract}

%
\section{Introduction}

One of the basic aims of research in highly relativistic nuclear collisions
has been the creation and detection of Quark Gluon Plasma (QGP). 
Such an endeavour started at the BNL AGS, continued at 
the CERN SPS and is presently being pursued at the heavy ion experiments 
at the BNL RHIC. The principal motivation behind such a sustained and
organized effort is that apart from being a confirmation of a long standing
prediction of thermal QCD, the production of deconfined matter would
actually recreate in the laboratory
conditions which existed in the microsecond universe. 
A variety of signals of this exotic state of matter have 
been proposed and studied extensively over the last two decades.
The suppression of the $J/\Psi$ particle, the enhanced production
of strange particles, specially strange antibaryons, excess 
production of photons and lepton pairs and
the formation of disoriented chiral condensates are some of the 
well accepted probes of a deconfinement and/or a chiral symmetry
restoring phase transition.
In the recently concluded Pb run at
the CERN SPS some of these, specially the magnitude of $J/\Psi$
suppression has been claimed to be inexplicable in a non-QGP
scenario~\cite{qm99}.

Owing to their long mean free paths, electromagnetic signals {\it viz.} 
photons and dileptons are by far the most direct probes of the 
collision~\cite{mt,gk,evs,ars,cb}. 
In fact they possess the unique property of
being sufficiently less prone to secondary interactions because of
their electromagnetic nature and yet are strong enough to
be experimentally detectable. Moreover, unlike hadronic signals, photons
and dileptons are produced all through the evolution process. In this 
connection one must bear in mind that
even if QGP is produced in the initial stages, the quarks and gluons 
will eventually form colour singlet hadrons before traveling to the detectors.
Hence, in order to unmask the emissions from the QGP one must
have a rather accurate estimation of the contributions from hadronic sources.
In this aspect, the dilepton spectrum enjoys a clear advantage.
This is because, though both photons and dileptons
couple to hadrons through spin one ($J^{P}=1^-$)
mesons, the dilepton spectra exhibits a resonant structure 
which, in the low mass regime includes the $\rho$ and 
the $\omega$ mesons.
Now, it has been emphasized that the properties of the vector mesons 
undergo nontrivial medium modifications in a hot and/or dense medium
such as likely to be produced in relativistic nuclear collisions.
Consequently the spectral modifications of these vector mesons
would be clearly revealed in the invariant mass spectra of the dileptons
through the shift of the resonant peaks.

Let us recall the various sources which produce dileptons in relativistic
nuclear collisions. 
The principal source of thermal dileptons from the QGP and hadronic phases 
are the quark-antiquark and pion annihilation processes respectively.
In the high mass region these compete with Drell-Yan pairs
which are produced in primary interactions between incoming partons in the very
early stages of the collision.
The $J/\Psi$ peak at 3 GeV marks the cut-off scale for 
thermal pairs from the plasma.
Around this region the decays of $D$ and $B$ mesons also become an important
source~\cite{gavin}.
 The vector mesons which decay both during the expansion and after
freeze-out can be identified from their characteristic peaks in the
spectrum. Below the vector mesons, Dalitz decays of
$\pi^0$, $\eta$, $\eta\prime$
and $\omega$ mesons provide the dominant source for dilepton production.

The observed enhancement of low-mass dileptons in Pb+Au collisions at
the CERN SPS compared to the yield from
hadronic decays at freeze-out 
as reported by the CERES Collaboration~\cite{ceres} 
has triggered a host of theoretical activities. Though the pion annihilation
channel $\pi^+\pi^- \ra l^+l^-$ accounts for a large 
fraction of
this enhancement, it turns out that a quantitative explanation of the data 
requires the incorporation of medium modifications of the vector mesons.
The enhancement of the dilepton yield in the low invariant mass
region was well reproduced by using reduced vector meson masses
in the medium~\cite{lkb,cass}.
Rapp {\it et al}~\cite{rcw}
have used a large broadening of the $\rho$ meson spectral function
due to scattering
off baryons in order to explain the data. 
However, both approaches rely
on a high baryon density for the dropping mass 
or the enlarged width of the $\rho$ meson.
Koch {\it et al}~\cite{koch} 
finds very little effect due to baryons 
on the dilepton yield in the kinematic range of CERES experiment. 
It is worth emphasizing here that as yet it
has not been possible to explain the observed low-mass enhancement
of dileptons measured in the Pb+Au collisions~\cite{cmhung}
as well as in the S-Au~\cite{solfrank}
collisions at the CERN SPS in a scenario which does {\it not} incorporate 
in-medium effects on the vector meson mass. 
In short,  the CERES dilepton data indicate a substantial change 
in the in-medium hadronic spectral function,
either mass shift or broadening but fails to make any distinction
between them.

We have shown earlier~\cite{wa98}
that the WA98 photon spectra can be 
explained by either of the two scenarios of relativistic nuclear collision:
(a) A\,+A\,$\ra$QGP$\ra$Mixed Phase$\ra$Hadronic Phase 
or (b) A\,+\,A\,$\ra$Hadronic Matter;
with downward shift of vector meson masses
and initial temperature $\sim 200$ MeV. The 
photon yield is insensitive to the broadening
of vector mesons~\cite{annals},
although the CERES dilepton data admits such a scenario
as mentioned above.
In the present work we would like to see whether 
the CERES dilepton data can be described by the
same scenarios ( a and b) and with similar initial conditions 
which reproduce the WA98 photon spectra. 
The effects of the thermal shift of the hadronic
spectral functions 
on both photon and dilepton emission have
been considered in Ref.~\cite{annals}
for an exhaustive set of models. However, an appreciable
change in the space-time integrated yield of electromagnetic
probes was observed for universal scaling 
and Quantum Hadrodynamic (QHD) model.
Therefore, in the present calculation we will
consider these two cases.

The paper is organized as follows. In the following section we will
briefly outline the phenomenology of medium effects on the vector mesons 
in the thermal environment which we have considered. In section~3 we will
consider the static dilepton rates due to different processes. Then in 
section~4 we will describe the dynamics of space-time evolution followed by the
results of our calculation in section~5. We will conclude with a
summary and discussions in section 6.

\section{Medium Effects}

A substantial amount of literature has been devoted  to the 
issue of temperature and/or density dependence of hadrons within
various models (see~\cite{annals,thpr,brpr,rapp,rdp} for a
review).
In Ref.~\cite{annals} we have discussed the effects of spectral 
changes of hadrons on the electromagnetic probes in detail.
In this work we will consider medium modifications of vector mesons
in two different scenarios where non-trivial effects on the dilepton 
spectra were seen. These are: the universal scaling hypothesis 
and the Quantum Hadrodynamic (QHD) model.
   
In the scaling hypothesis, the  parametrization of in-medium 
quantities (denoted by $*$) at finite temperature, $T$ and
baryon density, $n_B$~\cite{geb} is
\be
{m_{V}^* \over m_{V}}  = 
{f_{V}^* \over f_{V}} = 
{\omega_{0}^* \over \omega_{0}}  =\left(1-0.2\frac{n_B}{n_B^0}\right)
 \left( 1 - {T^2 \over T_c^2} \right) ^{\lambda},
\label{anst}
\ee
where $V$ stands for vector mesons, $f_V$ is 
the coupling between the electromagnetic current and the vector
meson field, $\omega_0$ is the continuum threshold,
$T_c$ is the critical temperature and $n_B^0$ is the
baryon density of normal nuclear matter.
Mass of the nucleon also varies with temperature
according to Eq.~(\ref{anst})
(pseudo scalar masses remain unchanged). 
The particular exponent $\lambda=1/2$ (1/6)
is known as Nambu (Brown-Rho) scaling. We will use $\lambda=1/2$ 
in our calculations.
Note that there is no definite reason to believe that all the in-medium
dynamical quantities are dictated by a single exponent $\lambda$.
This is the simplest possible ansatz (see~\cite{annals} for a discussion). 
The effective mass of $a_1$
is estimated by using Weinberg's sum rules~\cite{weinberg}.

In the Quantum Hadrodynamic model~\cite{vol16,chin} of nuclear matter
the vector meson properties are modified due to coupling with
nucleonic excitations. 
The nucleons interact through the exchange 
of scalar $\sigma$ and the vector $\omega$
mesons and their mass is modified due to the scalar condensate.
This is evaluated in the Relativistic Hartree Approximation (RHA).
Coupling with these modified nucleonic excitations induce
changes in the $\rho$ and $\omega$ meson masses.
This modification is contained in the meson self energy
which appears in the Dyson-Schwinger equation for the effective propagator
in the medium. The interaction vertices are provided by the Lagrangian
\be
{\cal L}^{\s int}_{VNN} = g_{VNN}\,\left({\bar N}\gamma_{\mu}
\tau^a N{V}_{a}^{\mu} - \frac{\kappa_V}{2M_N}{\bar N}
\sigma_{\mu \nu}\tau^a N\partial^{\nu}V_{a}^{\mu}\right),
\label{lagVNN}
\ee
where $V_a^{\mu} = \{\omega^{\mu},{\vec {\rho}}^{\mu}\}$,
$M_N$ is the free nucleon mass, $N$ is the nucleon field
and $\tau_a=\{1,{\vec {\tau}}\}$, $\vec\tau$ being the Pauli matrices.
The real part of the vector meson self-energy due to $N\bar N$ polarization
is responsible for the mass shift. The details of the calculations can be 
found in our previous work~\cite{npa1,npa2} (see also 
~\cite{csong}) and we do not reproduce them here.

\section{Dilepton Emission Rate}

In this section, we briefly recapitulate the main
equations relevant for evaluating dilepton emissions
from a thermal source.  
The emission rate of dileptons with four momentum $p_\mu$ 
can be expressed in terms of the imaginary part of
the retarded current-current correlation function 
as
\be
\frac{dR}{d^4p}=\frac{\alpha}{12\pi^4\,p^2}
{\s Im}W^R_{\mu\mu}\,f_{BE}(p_0)
\label{drcor1}
\ee
where the electron mass has been taken to be zero.
Putting $d^4p=\pi\,dM^2\,p_Tdp_T\,dy$, where $M$, $p_T$ and $y$
are the invariant mass, transverse momentum and rapidity of the
lepton pair respectively, 
this is written as
\be
\frac{dR}{dM^2\,p_T\,dp_T\,dy}=\frac{\alpha}{12\pi^3}\frac{1}{M^2}
\,e^{-M_T\cosh y/T}
\,\,{\s Im}W^{R}_{\mu\mu}.
\label{drcor_bol}
\ee
in the Boltzmann approximation.

The information about the specific processes occurring in the thermal
medium which produces the dileptons resides in the current 
correlator $W^{\mn}$.
In the QGP, the dominant channel for dileptons production is
quark-antiquark annihilation. 
The rate for this process is obtained from the lowest order
diagram contributing to the current correlator $W^{\mn}$.
In the limit of vanishing quark masses,
the rate for the process $q\bar q\,\ra\,e^+e^-$ is obtained as,
\be
\frac{dR}{dM^2\,p_T\,dp_T\,dy}=\frac{5}{9}\,\frac{\alpha^2}{4\pi^3}
\,e^{-M_T\cosh y/T}.
\label{qqbar_bol}
\ee

Now let us consider dilepton production in the hadronic medium.
The process $\pi^+\pi^-\,\ra\,e^+e^-$ is known to be the most dominant
source of dilepton production from hadronic matter. 
In order to obtain the rate of dilepton emission due to pion annihilation
from Eq.~(\ref{drcor_bol})
the electromagnetic current correlator is expressed in terms of
the effective propagator of the vector particle in the thermal medium
using vector meson dominance (VMD) so as to obtain
\be
{\s Im}W_{\mu\mu}^R=\frac{3e^2 m_\rho^{\ast 4}}{g_{\rpp}^2}
\left[\frac{{\s Im}\Pi_{\rho}^R}{(M^2-m_\rho^{\ast 2})^2
+[{\s Im}\Pi_{\rho}^R]^2}\right].
\label{walcor}
\ee
where ${\s{Im}}\Pi_{\rho}^R$ is the self energy of the $\rho$ arising out of 
interaction with excitations in the medium and $m_\rho^{\ast 2}
=m_\rho^2+({\s Re}\Pi_\rho^{R})^2$. 
For a $\rho$ meson propagating
with energy $p_0$ and three momentum $\vec p$ the width is given by  
\bea
\Gamma_{\rho\,\rightarrow\,\pi\,\pi}(p_0, {\vec p}) 
&=& {g_{\rho \pi \pi}^2 \over 48\pi}\,W^3(s)\,\frac{s}{p_0}\,
\left[1+\frac{2T}{W(s)\sqrt{p_0^2-s}}\right.\nonumber\\
&&\times\left.\ln\left\{\frac{1-\exp[-\frac{\beta}{2}(p_0+W(s)
\sqrt{p_0^2-s})]}
{1-\exp[-\frac{\beta}{2}(p_0-W(s)\sqrt{p_0^2-s})]}\right\}
\right]
\eea
where $s = M^2 = p_0^2-{\vec p}^2$ and $W(s) = \sqrt{1-4m_{\pi}^2/s}$.
It can be easily shown that this expression reduces to the usual expression 
for the rho width in the limit $\vec p\ra 0$,
\be
\Gamma_{\rho\,\rightarrow\,\pi\,\pi}(p_0) = 
\frac{g_{\rho\,\pi\,\pi}^2}{48\pi}\,p_0 \,W^3(p_0)
\left[\left(1+f_{BE}(\frac{p_0}{2})\right)\,\left(1+f_{BE}(\frac{
p_0}{2})\right)-f_{BE}(\frac{p_0}{2})f_{BE}(\frac{p_0}{2})
\right]
\label{width}
\ee
with $f_{BE}(x) = [e^x -1]^{-1}$.

We had shown earlier~\cite{chrono} that the width of the $\omega$ meson 
can be large due to various reactions occurring in the
thermal bath. The most dominant process, among others,
is $\omega\pi\,\ra\,\pi\pi$~\cite{asrds}. In view of this
we also take into account the in-medium decay $\omega\,\ra\,e^+e^-$.
Weldon~\cite{weldon93} had shown that the production
of an {\em off-shell} vector meson ($V$)  of four momentum 
$p$ (where $p^2=M^2$) in a thermal medium and its subsequent
decay into a lepton pair leads to the dilepton emission rate
\be
\frac{dR}{d^4p}=2\frac{(2J+1)}{(2\pi)^3}\,f_{BE}\,M\Gamma^{\s vac}_{V\,\ra\,l^+l^-}\,
\left[\frac{1}{\pi}\frac{{\s Im}\Pi_V^R}{(p^2-m_V^{\ast\,2})^2+({\s Im}\Pi_V^R)^2}
\right].
\label{wel1}
\ee
As before this reduces to
\be
\frac{dR}{dM^2\,p_T\,dp_T\,dy}=\frac{3}{4\pi^3}
\Gamma^{\s vac}_{V\,\ra\,l^+l^-}
\frac{M^2\,\Gamma_V(M)}
{(M^2-m_V^{\ast 2})^2
+M^2\Gamma_V^2(M)}
\,e^{-M_T\cosh y/T}
\label{decay_bol}
\ee
with the vacuum decay width~\cite{lkb,cass} given by
\be
\Gamma_{V\,\ra\,l^+\,l^-}^{{\s vac}}=
\frac{4\pi\alpha^2}{3g_V^2}\frac{m_V{^*4}}{M^3}.
\ee

\section{Space-Time Evolution}

The observed spectrum originating from an expanding 
hadronic matter is obtained by convoluting the static
rates given above with the expansion dynamics. The usual
space-time picture of the evolution is as follows.
Matter initially formed as a
hot hadronic gas expands and cools till freeze-out. 
However, in a phase transition scenario, matter produced 
as QGP expands and cools to the phase transition temperature $T_c$. A
mixed phase of coexisting quark and hadronic matter follows. After all the
quark matter has converted to hadrons, the hot hadron matter expands and
cools till freeze-out. 
In this work we use Bjorken-like~\cite{bjorken}
hydrodynamical model for the isentropic expansion of the matter.
The essential input at this stage is
the equation of state (EOS) which provides the cooling law.
For the QGP sector we use the bag model equation of state with
two flavour degrees of freedom. The temperature in the QGP phase evolves
according to Bjorken scaling law $T^3\,\tau=T_i^3\tau_i$.

The hadronic phase is taken to consist of $\pi$, $\rho$, $\omega$, 
$\eta$ and $a_1$ 
mesons and nucleons.
The energy density and pressure
for such a system is given by,
\be
\epsilon_H=\sum_{i=mesons} \frac{g_i}{(2\pi)^3} 
\int d^3p\,E_i\,f_{BE}(E_i,T)
+\frac{g_N}{(2\pi)^3} 
\int d^3p\,E_N\,f_{FD}(E_N,T)
\ee
and
\be
P_H=\sum_{i=mesons} \frac{g_i}{(2\pi)^3} 
\int d^3p\frac{p^2}{3\,E_i}f_{BE}(E_i,T)
+\frac{g_N}{(2\pi)^3} 
\int d^3p\frac{p^2}{3\,E_N}f_{FD}(E_N,T)
\ee
where the sum is over all the mesons under consideration and $N$ stands
for nucleons and $E_i=\sqrt{p^2 + m_i^2}$.          
The entropy density is parametrized as,
\be
s_H=\frac{\epsilon_H+P_H}{T}\,\equiv\,4a_{\s{eff}}(T)\,T^3
= 4\frac{\pi^2}{90} g_{\s{eff}}(m^\ast,T)T^3
\label{entro}
\ee
where  $g_{\s{eff}}$ is the effective statistical degeneracy.
Thus, we can visualize the finite mass of the hadrons
having an effective degeneracy $g_{\s{eff}}(m^\ast,T)$. 
(For a possible parametrization of entropy density consistent
with lattice QCD we refer to ~\cite{asakawa}). We will see later
that the value of $g_{\s {eff}}$ is large at high temperature.
The variation 
of temperature from its initial value  $T_i$ to final value 
$T_f$ (freeze-out temperature) with proper time ($\tau$) is governed 
by the entropy conservation 
\be
s(T)\tau=s(T_i)\tau_i
\label{entro1}
\ee
Similarly, the evolution of the baryon density is governed by the 
conservation of baryonic current.
In the present scenario 
the velocity of sound, $c_s$ becomes a function of $T$,
\be
c_s^{-2}=\frac{T}{s_H}\frac{ds_H}{dT}=
\left[\frac{T}{g_{\s {eff}}}\frac{dg_{\s {eff}}}{dT}+3\right]
\label{cs}
\ee
We will see later that $c_s$ differs substantially from its value
corresponding to an ideal gas. We have considered a first
order phase transition scenario where the the entropy
density in the mixed phase is given by 
\be
s_{mix}=f_Q\,s_{Q}(T_c)\,+\,(1-f_{Q})\,s_H(T_c)
\ee
where $s_{Q}$ ($s_H$)is the entropy density in
the QGP (hadronic) phase at temperature $T_c$
and the function $f_Q$ ($1-f_Q$) is the
volume fraction of QGP (hadronic) phase,
obtained by solving the equation which
governs the isentropic expansion of
the matter.
The initial temperature of the system is obtained by solving
the following equation
\be
\frac{dN_\pi}{dy}=\frac{45\zeta(3)}{2\pi^4}\pi\,R_A^2 4a_{\s{eff}}T_i^3\tau_i
\label{dndy}
\ee
where $dN_\pi/dy$ is the total pion multiplicity
($\sim 1.5\times$ charge multiplicity), $R_A$ is the radius
of the system, $\tau_i$ is the initial thermalization time and 
$a_{\s{eff}}=({\pi^2}/{90})\,g_{\s{eff}}(T_i)$. 
When the system is produced in the QGP phase, $g_{\s{eff}}$ in the
above equation is replaced by $g_{QGP}$ which 
for two quark flavours is 37. If the quark and gluon 
masses are non-zero in the thermal bath then the effective
degeneracy of the QGP, $g_{\s{QGP}}^{\s{eff}}$, defined by Eq.~(\ref{entro})
can be lower than 37, resulting in a higher value of $T_i$ for
a given multiplicity according to Eq.~(\ref{dndy}).
Note that the change in the expansion dynamics
as well as the value of the initial temperature due
to medium effects relevant for a hot hadronic gas
also enters through the effective statistical degeneracy.

\section{Results}

We will now compare 
our results with experimental spectra obtained by the CERES collaboration
in Pb+Au collisions at the CERN SPS. 
In the central
rapidity region the entropy per baryon is 
found to be quite large $\sim 40-50$~\cite{pbm,dumitru},
consequently the effects of finite baryon density
will be small. We have confirmed this by 
assuming an initial baryon density of
$n_B^i=2.5n_B^0$ for $dN_{ch}/d\eta=150,\, 210,\,270,\,$ and
$n_B^i=4.5n_B^0$ for $dN_{ch}/d\eta=350,\, where $ $n_B^0$
is the baryon density of normal nuclear matter. The baryonic
contribution is further suppressed due to the limited
acceptance of the CERES detector~\cite{MB,PH}(see also~\cite{RAS}).
The dilepton spectra for various charge multiplicities 
are evaluated both with and without a phase transition. 
In all the cases we will assume an initial formation time $\tau_i= 1$ fm/c
and freeze-out temperature, $T_f=120$ MeV~\cite{na49}.
The initial temperatures depend on the scenario concerned.
They are listed below in Table~1. The second column gives the initial 
temperatures in a phase transition scenario with QGP as the initial state.
The next two columns are the values corresponding to
a hot hadronic gas scenario with vacuum and QHD masses respectively.
When mass modifications are incorporated using
universal scaling the initial state depends on the particular value of $T_c$
that one considers. In view of the prevailing uncertainties in its 
value we have considered $T_c=$170, 190 and 200 MeV~\cite{karsch}.
For $dN_{ch}/d\eta=150$ and 210, $T_i\leq 170$ MeV. Therefore,
for these cases we consider the hadronic initial states only. 
However, for events with $dN_{ch}/d\eta$=270 and 350
the initial temperature can be larger than the critical
temperature as seen below.

\renewcommand{\arraystretch}{1.5}
\vskip 0.2in
\begin{center}
\begin{tabular}{|c|c|c|c|c|c|}
\hline
& QGP & vacuum & QHD & Scaling         & Scaling \\
&     & mass   &     & $T_c$=170 MeV & $T_c$=200 MeV \\
\hline
$dN_{ch}/d\eta$ & $T_i$ (MeV) & $T_i$ (MeV) & $T_i$ (MeV) & $T_i$ (MeV) & $T_i$ (MeV) \\
\hline
150 & 161 & 209 & 199 & 168 & 181 \\
210 & 171 & 219 & 206 & 171 & 187 \\
270 & 179 & 227 & 212 & 179 & 192 \\
350 & 187 & 235 & 217 & 187 & 196 \\
\hline
\end{tabular}
\end{center}
Table 1 : Values of initial temperature 
obtained by solving Eqs.~\ref{entro}
and ~\ref{dndy} self-consistently
for various values of $dN_{ch}/d\eta$. 
\vskip 0.1in

The velocity of sound, which governs the space-time
evolution of the system,  is obtained by solving 
Eqs.~(\ref{entro}) and (\ref{cs}) self consistently. 
This is shown in Fig.~\ref{cs2}
where hadronic masses vary
as a function of temperature according
to  universal scaling and QHD model.  
It is noted that the value of $c_s^2$ in the hadronic phase 
is less than that of a massless ideal gas ($c_s^{ideal}=1/\sqrt{3}$)
indicating substantial interaction among the various constituents
of the system. This is consistent with the results of 
Ref.~\cite{asakawa}.

\bef
\centerline{\psfig{figure=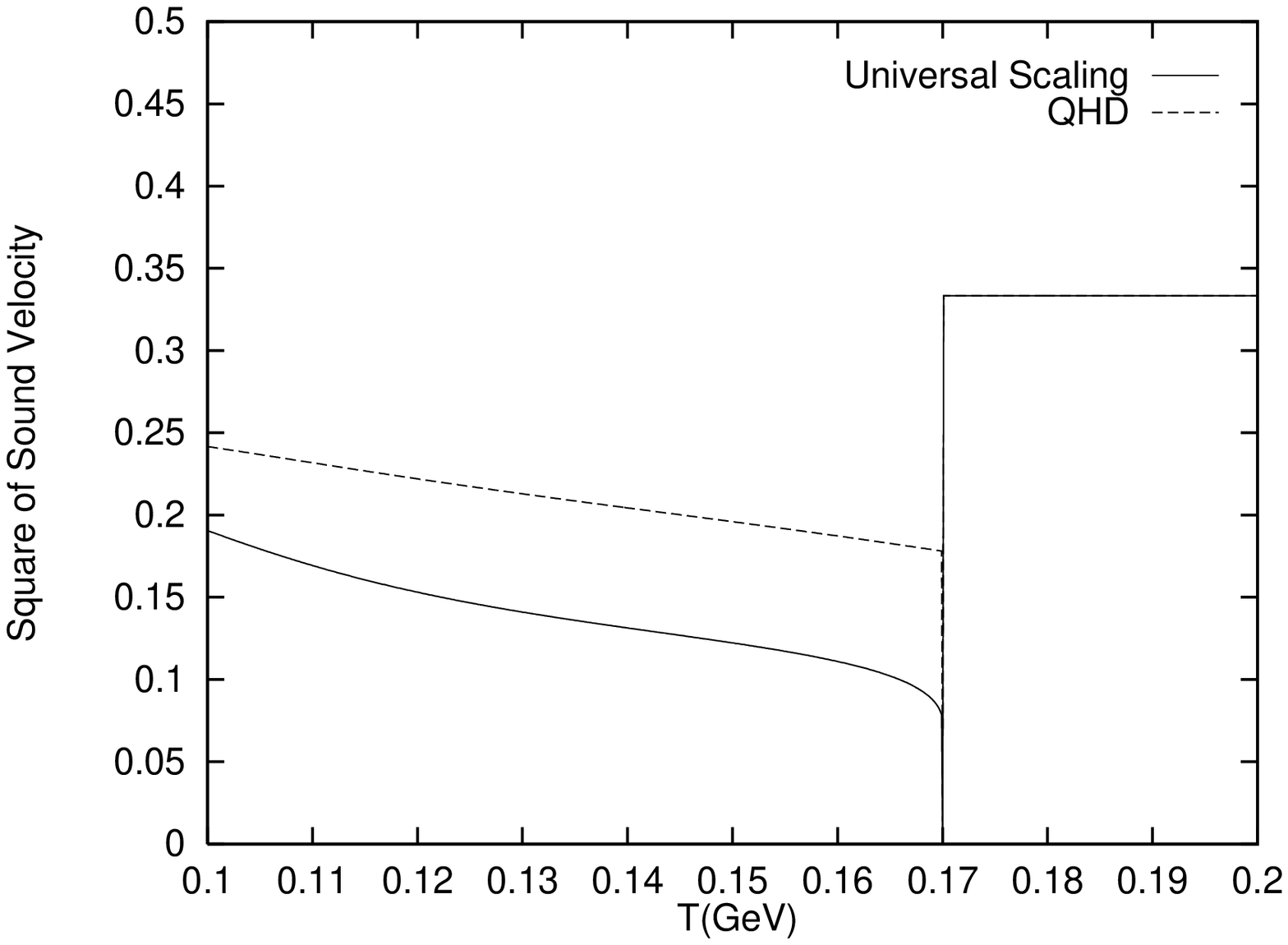,height=6cm,width=8cm}}
\caption{Velocity of sound as a function of temperature,
for different mass variation scenario for the hadrons
in a deconfinement phase transition scenario.
}
\label{cs2}
\eef

In order to substantiate the point further, we now show in Fig.~\ref{deg}
how the effective degeneracy of the hadronic matter varies with temperature.
The quantity $g_{\s{eff}}$ as obtained from Eq.~(\ref{entro})  
contains the effect of interactions among the various hadronic constituents
which gives rise to the medium modified masses. We observe that when the masses
vary according to Nambu scaling law, the value of $g_{\s{eff}}$ nearly 
approaches 30 which is more than thrice the value one obtains when vacuum
masses are considered. As emphasized earlier, this variation has non-trivial 
effects on the cooling law of the system described by Eq.~(\ref{entro1})
and is displayed in Fig.~\ref{ttau}.
\bef
\centerline{\psfig{figure=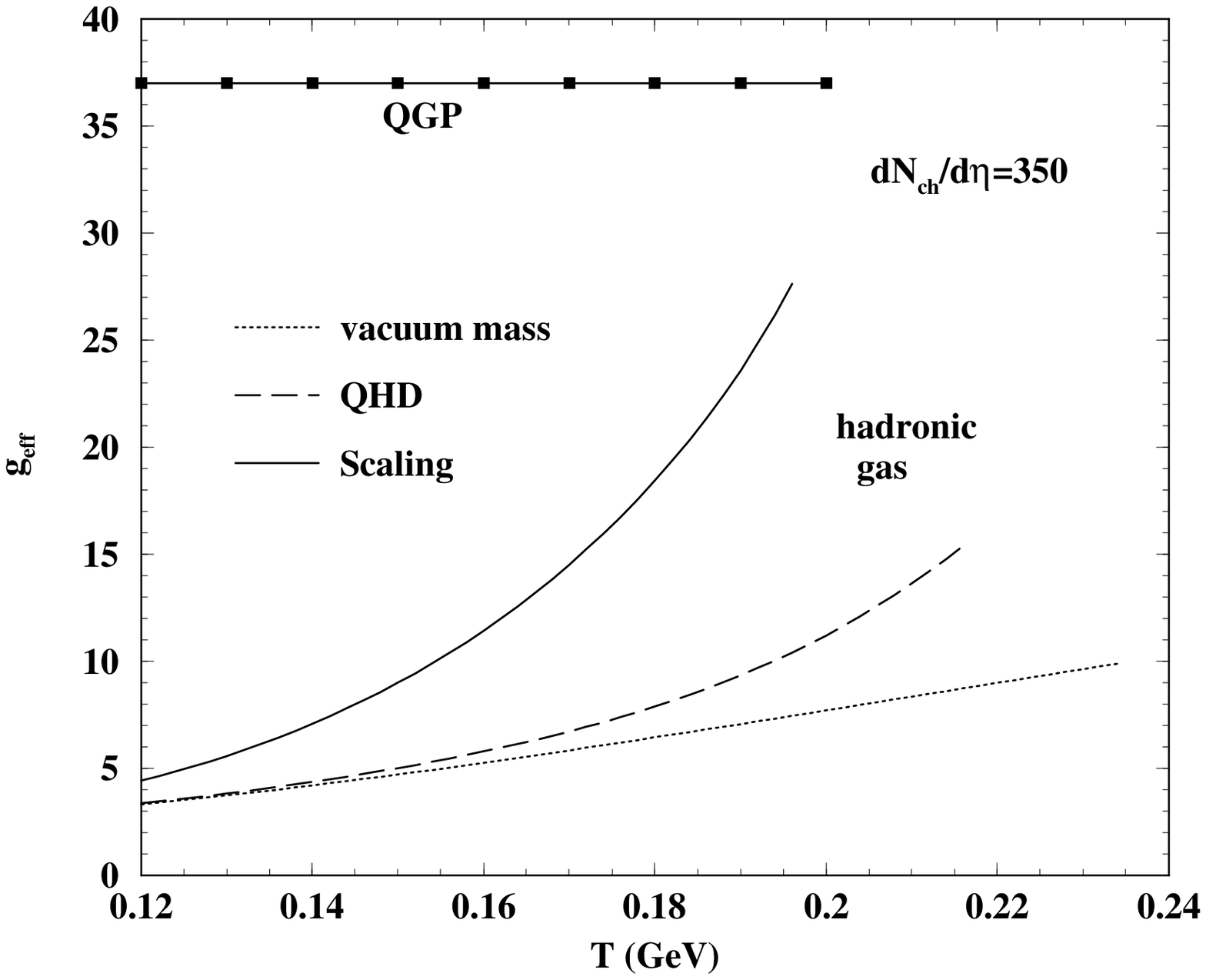,height=6cm,width=8cm}}
\caption{The effective degeneracy as a function of temperature}
\label{deg}
\eef
\bef
\centerline{\psfig{figure=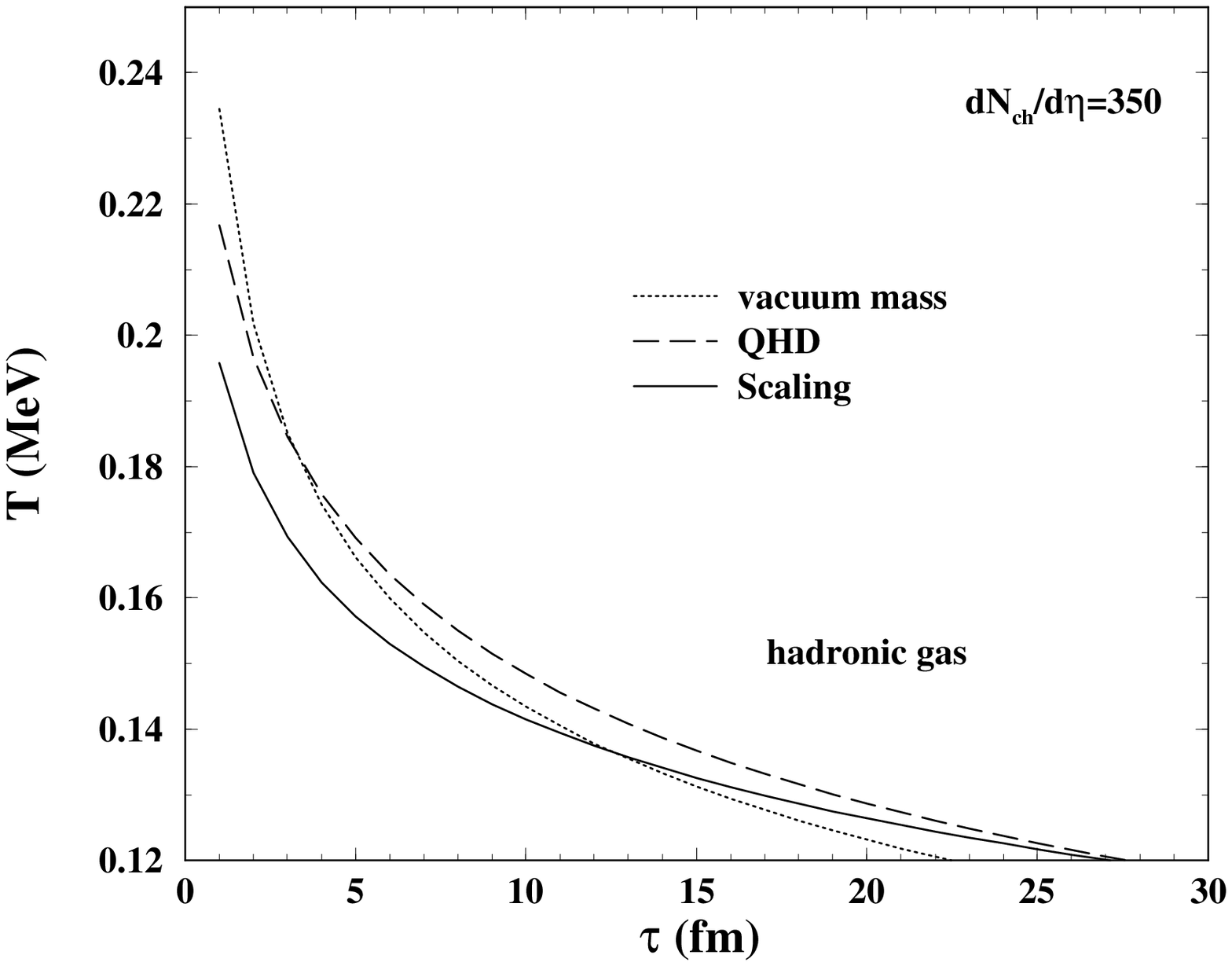,height=6cm,width=8cm}}
\caption{The temperature as a function of time}
\label{ttau}
\eef

We have considered dileptons with transverse 
momentum $p_T$ above 200 MeV/c
and an opening angle $\Theta_{ee}>$ 35 mrad.
These are kinematical cuts relevant for the CERES detector 
and are incorporated as described in ~\cite{solfrank}. 
In all the figures, the quantity $\lgl N_{ch}\rgl$ 
indicates the average number
of charged particles per unit rapidity interval in the pseudorapidity
range $2.1<\eta<2.65$. In all the cases background due to hadron
decays are added to the thermal yields.

In Figs.~\ref{f150} and \ref{f210} we have shown the invariant mass 
spectra of
dileptons corresponding to $\lgl N_{ch}\rgl$=150 and 210 respectively.
In these cases the initial temperature is not high enough to
favour a phase transition to QGP.
\bef
\centerline{\psfig{figure=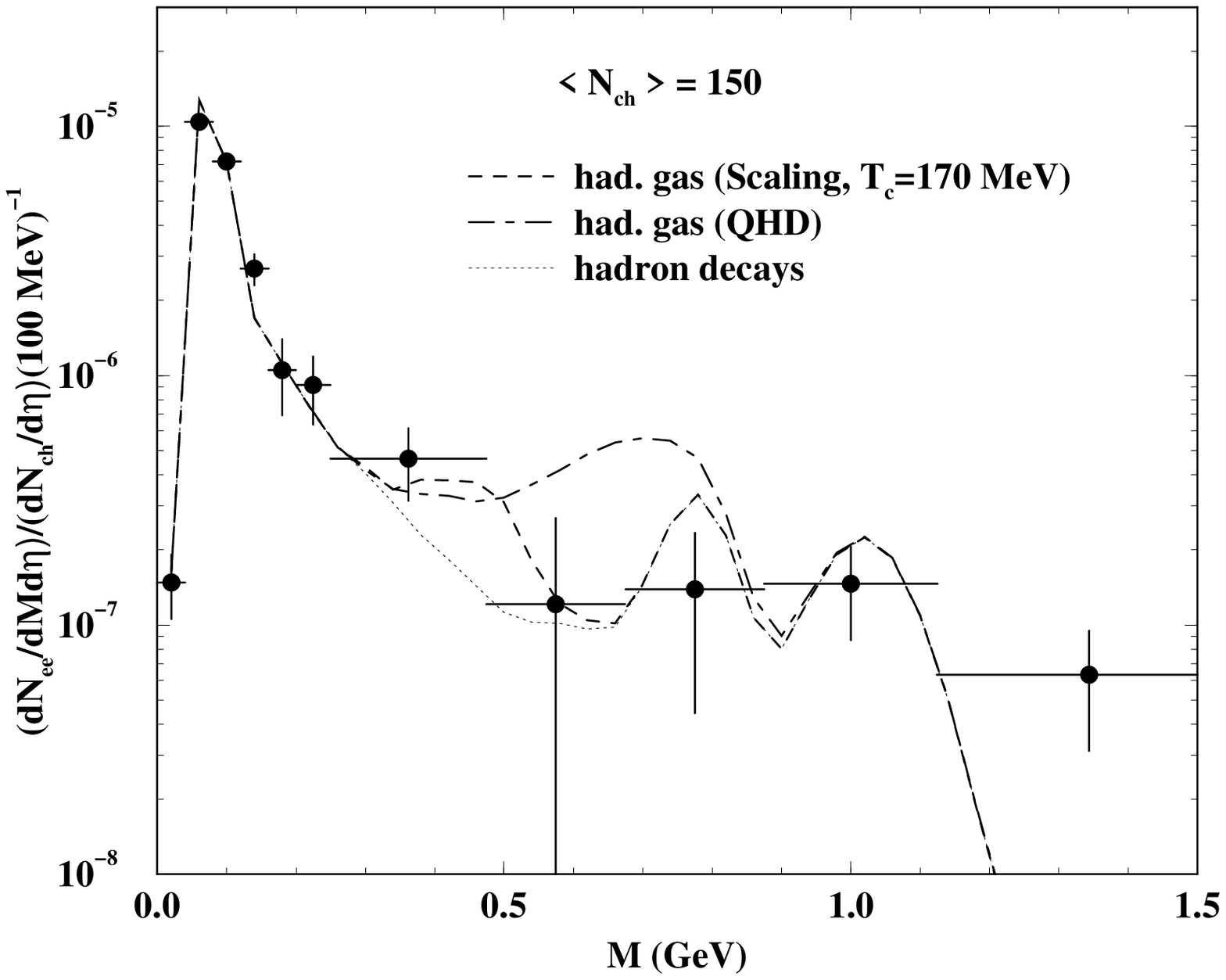,height=6cm,width=8cm}}
\caption{Dilepton spectra for $\lgl N_{ch}\rgl$=150. 
The dashed line (dash-dotted) indicates $e^+e^-$ 
pair yield when the system is formed in the hadronic phase and
hadronic masses vary according to
universal scaling (QHD). Note that these also include the yield from hadronic
decays at freezeout (dotted line). 
}
\label{f150}
\eef
Hence we have considered a hot hadronic gas as the initial state. The medium
effects are considered according to universal scaling (Eq.~(\ref{anst}) with 
$\lambda=1/2$) and the
QHD model. For $dN_{ch}/d\eta=150$ a good description of the
data is realized with $T_i=168$ MeV obtained by solving
Eq.~(\ref{dndy}) for the mass variation
given in Eq.~(\ref{anst}) with $\lambda=1/2$ 
(dashed line). For QHD
however, $T_i=199$ MeV (because of lower $g_{\s{eff}}$)
resulting
in a larger dilepton yield. The theoretical yield (dash-dotted 
line) overestimate the
data in the invariant mass region, $M\sim 0.5$ GeV
because of smaller reduction in $\rho$-mass in QHD.
For $dN_{ch}/d\eta=210$ the initial temperatures
are $T_i=171$ and $206$ MeV for universal
scaling and QHD model respectively. The dilepton yield
is well reproduced for both the cases. Considering the
experimental error bars it is not possible 
to state which one of the two is compatible with the data.

In Fig.~\ref{f270} we display the results for $dN_{ch}/d\eta=270$.
In this case apart from the hadronic gas 
scenario we have also considered a deconfined initial state with
 $T_i=179$ MeV 
which
evolves into a hadronic gas via a mixed phase.  
The observed enhancement of the dilepton yield
around $M\sim 0.3 - 0.6 $ GeV can be reproduced
with the QGP initial state, once the variation
of vector meson masses in the mixed and the hadronic phases
are taken into account (solid line).
The data is also reproduced 
by a hadronic initial state with $T_i=212 $ and $192$
MeV for QHD (dash-dotted) and universal mass variation scenario (dashed)
respectively. The $\rho$-peak in the dilepton spectra shifts towards
lower $M$ for universal scaling compared to QHD model, indicating
a strong medium effect in the former case. In this case again it
is impossible to differentiate between the QHD model and 
the universal scaling scenario.

\bef
\centerline{\psfig{figure=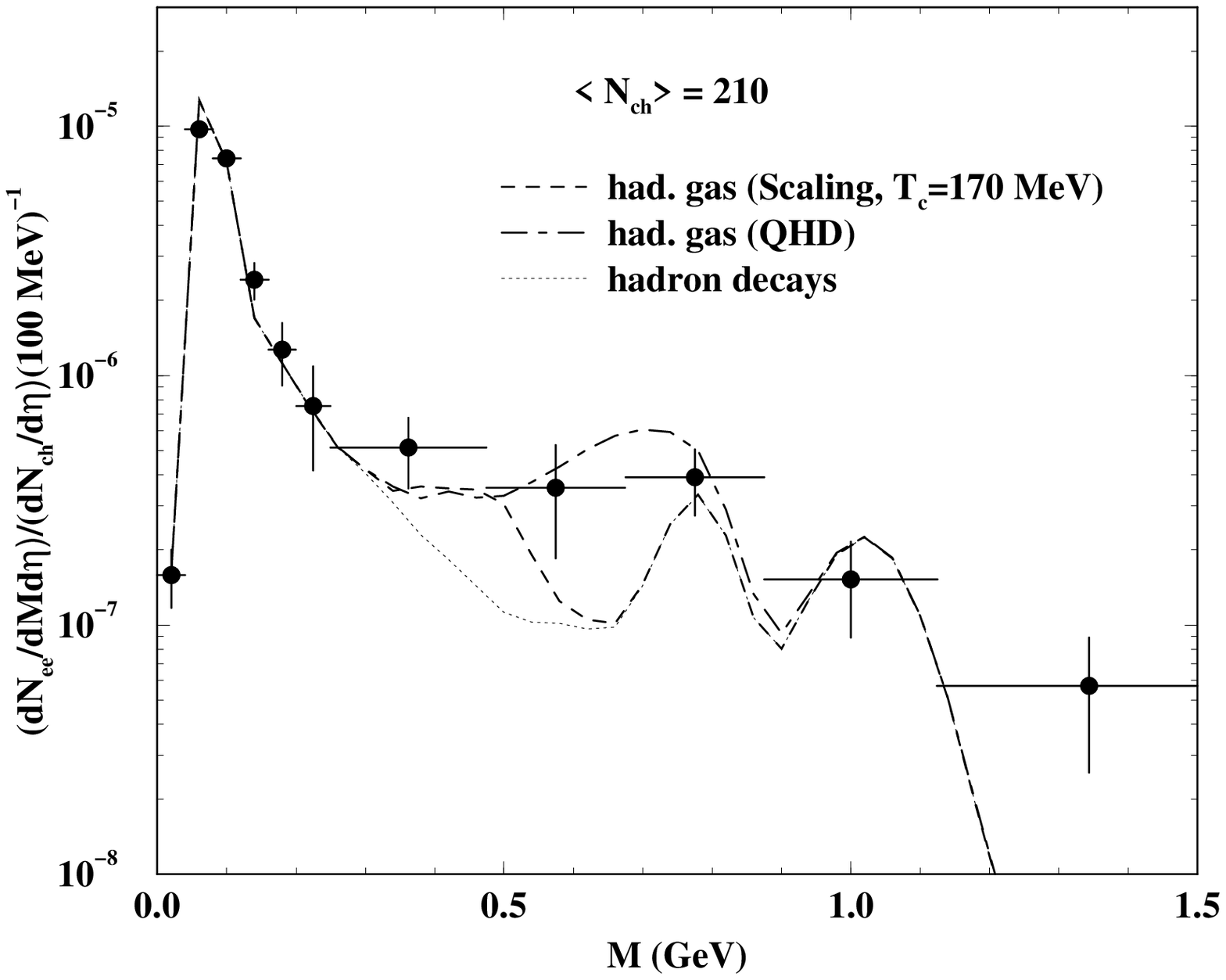,height=6cm,width=8cm}}
\caption{Same as Fig.~\protect\ref{f150} for $\lgl N_{ch}\rgl$=210.
}
\label{f210}
\eef

In Fig.~\ref{f350} we depict the dilepton spectra
for QGP and hadronic initial states for $dN_{ch}/d\eta=350$. 
Results with
hadronic initial state and universal scaling of
hadronic masses with temperature seem to describe the data
reasonably well. 
We have also noted that with the temperature dependent mass from QHD  
model can not reproduce the low $M$ enhancement of the experimental
yield.  
For $T_i=187$ MeV (see table 1) with a QGP initial state
it is not possible to explain the measured dilepton spectra
(result not shown in the figure).  
However, as we mentioned in section 4, 
the value of $T_i$ in the 2-flavour QGP may be larger
if $g_{QGP}$ is lower than 37 due to interactions
among quarks and gluons (leading to non-zero
masses of quarks and gluons) in the QGP phase.
A good description of the data 
can be obtained by taking $T_i=200$ MeV 
with QGP initial state for $T_c=190$ MeV.  
All these results seem to indicate an initial temperature
$\sim 200$ MeV, a value which we obtained earlier by analyzing WA98
photon data. We also show the results due to large broadening
of the $\rho$ spectral function in the medium. The broadening
of $\rho$ is modelled by assuming the temperature dependent
width as : $\Gamma_\rho(T)=\Gamma_\rho(0)/(1-T^2/T_c^2)$.
 
\bef
\centerline{\psfig{figure=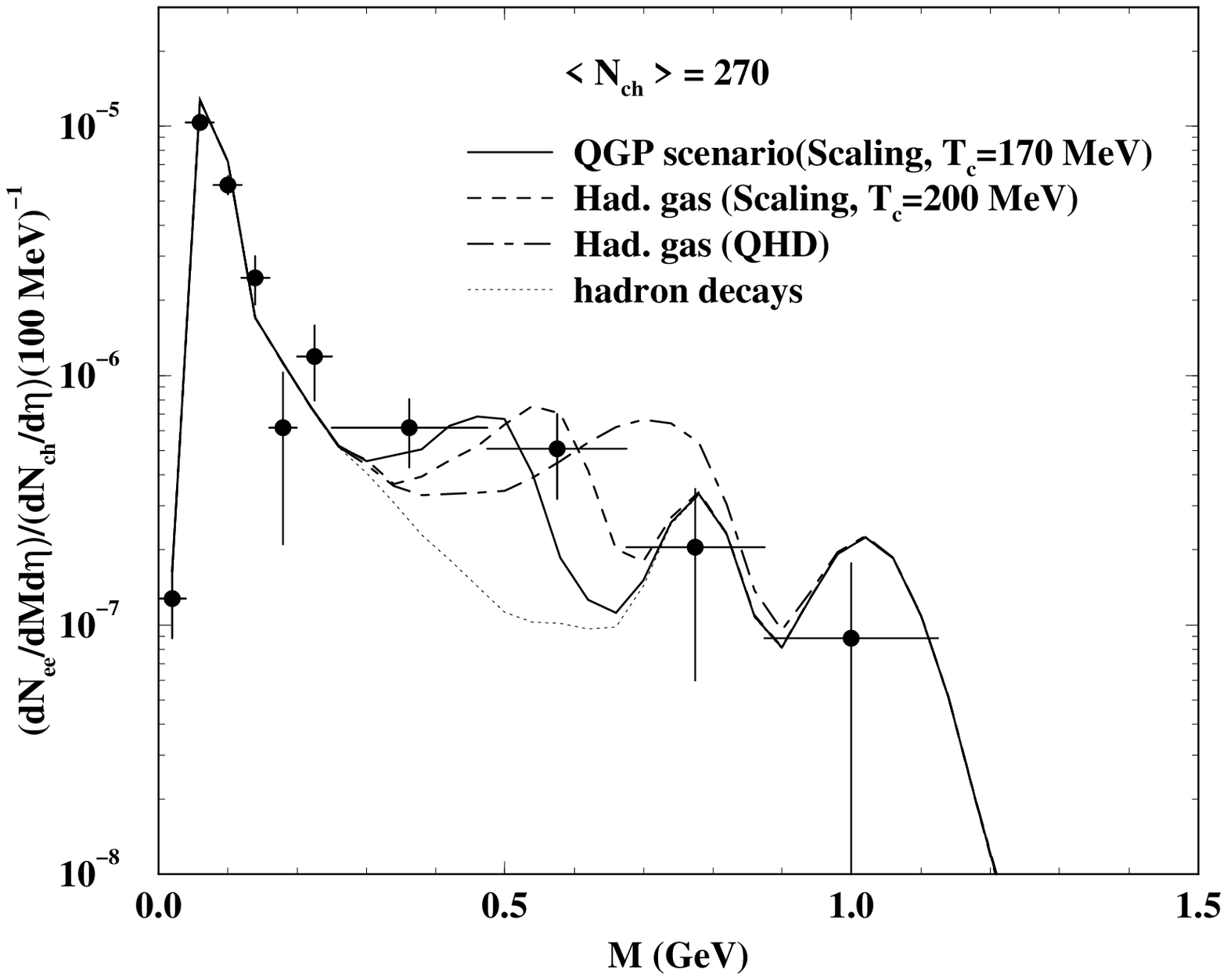,height=6cm,width=8cm}}
\caption{Dilepton spectra for $\lgl N_{ch}\rgl$=270 for
different initial states and mass variation scenarios.
Solid line indicates dilepton yield when
QGP is formed. (Here
QGP scenario indicates sum of yields from 
QGP phase+QGP part of the mixed phase 
+hadronic part of the mixed phase +hadronic phase)
The dashed line (dash-dotted) indicates $e^+e^-$ 
pair yield when the system is formed in the hadronic phase and
vector meson masses vary according to
universal scaling (QHD).  
}
\label{f270}
\eef

Finally, in Fig.~\ref{phot_ro} we demonstrate the effects of a large
broadening of the vector meson spectral function on the photon spectra.
Here we are interested in the relative shift of the photon
spectra with vacuum mass and decay width of $\rho$ in comparison
to the spectra obtained with vacuum masses but large broadening
of $\rho$.
To estimate the photon yield from hadronic matter, we have
considered the reactions~\cite{kapusta},
$\pi\,\rho\,\ra\, \pi\,\gamma$, 
$\pi\,\pi\,\ra\, \rho\,\gamma$, $\pi\,\pi\,\ra\, \eta\,\gamma$, 
$\pi\,\eta\,\ra\, \pi\,\gamma$ and the decays $\rho\,\ra\,\pi\,\pi\,\gamma$
and $\omega\,\ra\,\pi\,\gamma$.
The invariant amplitudes for all these 
processes can be found in Refs.~\cite{npa1,npa2}.
We have  also considered photon emission
due to the reaction $\pi\,\rho\,\ra\,a_1\,\ra\,\pi\,\gamma
~\cite{annals}$. 
The width of $\rho$ has been taken as 1 GeV, independent
of temperature purely for the purpose of demonstration. 
In a thermal medium, even if a huge enhancement
of the width takes place as a result of interactions, near the freeze-out
temperature the vector meson will regain its vacuum properties.
So, by taking a large constant width throughout the
evolution process we have overestimated its effect
on the photon spectra. However, it is clear from Fig.~\ref{phot_ro},
the effect is still small; the photon data with the current
statistics can not detect such effects. 

Throughout this work we have assumed thermal equilibrium
of the system. How good is this assumption? We recall
that the number of $\pi-\pi$ collisions, $N_{\pi\pi}$
in the region $M\sim m_{\rho}^*$ can be estimated from the
following equation~\cite{kkmr}:
\be
N_{\pi\pi}=2\times 3\times\frac{\Gamma_{\rho}^*}{B_{\rho}}
\frac{dN}{dMd\eta}|_{M=m_{\rho}^*}
\ee
where $B_{\rho}$ is the branching ratio for 
the decay $\rho\,\ra\,e^+e^-$. The factor 3 comes from
the isospin combinations and counting collisions per 
particle gives a further factor of 2.
We assume that 
thermal equilibrium may be realized in the 
system if $N_{\pi\pi}\,\ge\,1$ (a similar argument
was put forward in ref.~\cite{shuryak} regarding the equilibration of gluons).
Considering the total multiplicity $=1.5\,\times\,dN_{ch}/d\eta$,
we point out that thermal equilibrium may
not be realized in the case of 
$dN_{ch}/d\eta=150$ and 210 since the number of
collisions per particle in the system,
$\sim N_{\pi\pi}/(1.5dN_{ch}/d\eta) < 1$.
Interestingly, we note that for $dN_{ch}/d\eta=150$
the thermal source is not required to describe the data; 
the yield from hadron decays explain the data well.
For $dN_{ch}/d\eta=210$, however, the yield from hadronic decays underestimate
the data.
For higher multiplicities ($dN_{ch}/d\eta=270$ and 350)
the number of collisions per particle
$\sim 2 - 3$. This is because for higher multiplicities ($<N_{ch}>$
=270, 350) the pion density ($\sim T^3$) increases by a factor of 1.6 and
the number of $\pi\,-\pi\,$ collisions is proportional 
to the square of the pion density.
This indicates that thermal equilibrium may have been achieved, 
justifying the use of hydrodynamics
to describe the space time evolution of the
system. 

We have also checked that the condition,
$\tau_{\s{scatt}}^{\pi}<\tau_{\s{exp}}$ is satisfied throughout
the evolution process, indicating that the use of hydrodynamics
is reasonable for space-time description~\cite{navara}. Here
$\tau_{\s{scatt}}^j=(\sum_i\langle v_{ij}\rangle\sigma_{ij}n_j)^{-1}$,
is the mean collision time, $v_{ij}$ is the 
relative velocity, $\sigma_{ij}$ is the scattering cross section,
$n_j$ is the particle density of specie $j$
and $\tau_{\s{exp}}=\tau_i(T_i/T)^{1/c_s^2}$ is the expansion time scale. 
\bef
\centerline{\psfig{figure=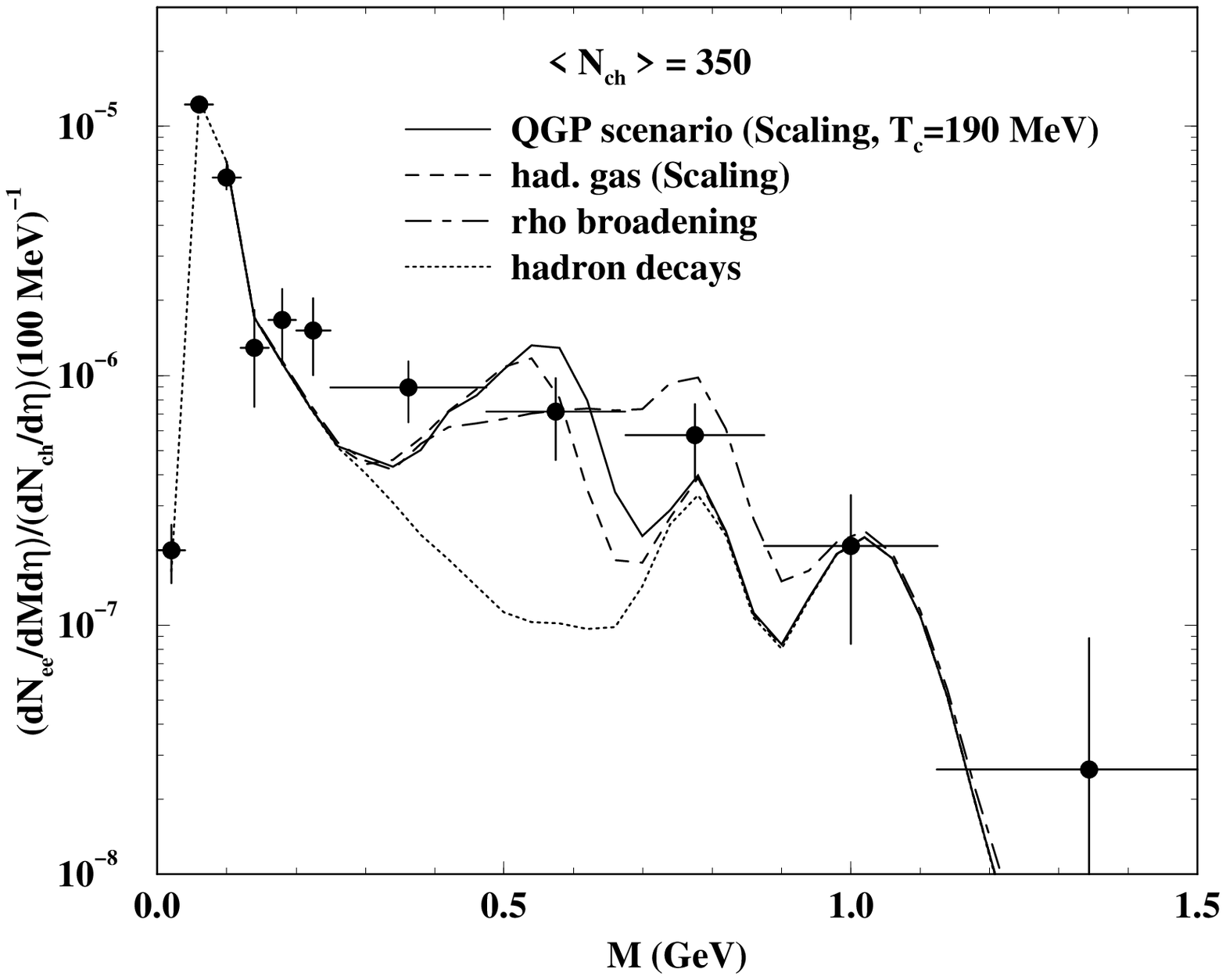,height=6cm,width=8cm}}
\caption{Dilepton spectra for $\lgl N_{ch}\rgl$=350 for
different initial states and mass variation scenarios.
Solid line indicates dilepton yield when
QGP is formed with $T_i=200$ MeV and $T_c=190$ MeV. 
The dash-dotted line indicates $e^+e^-$ 
pair yield with $\rho$ broadening and dashed line represents
results when the system is formed in the hadronic phase and
vector meson masses vary according to
universal scaling. The dotted line shows dilepton yield from hadronic decays. 
}
\label{f350}
\eef

\bef
\centerline{\psfig{figure=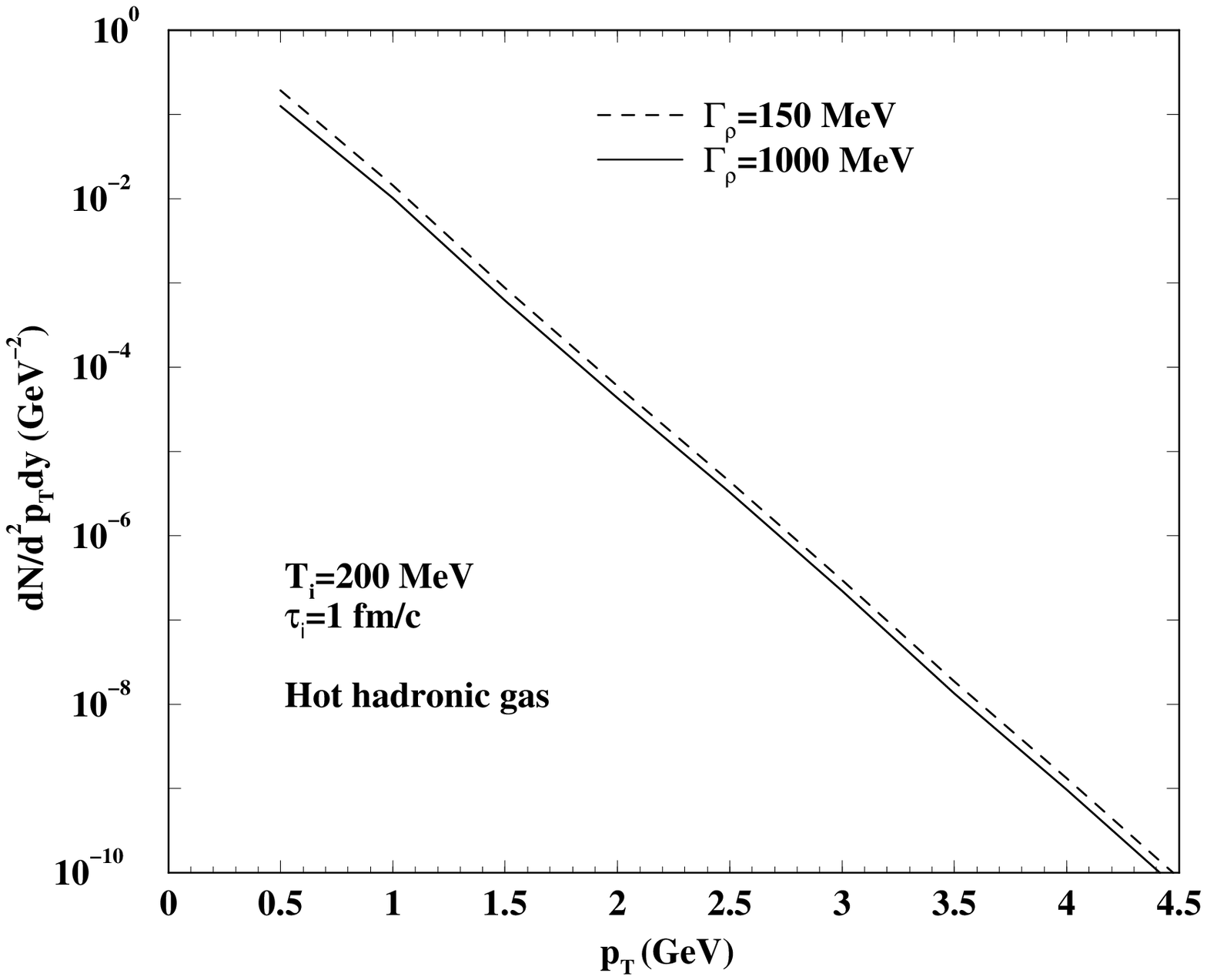,height=6cm,width=8cm}}
\caption{Thermal photon spectra for $T_i=200$ MeV and 
$\tau_i=1$ fm/c. 
Solid (dashed) line indicates photon yield with large
broadening of the $\rho$ spectral function (vacuum spectral
function). 
}
\label{phot_ro}
\eef

It was proposed earlier ~\cite{kkmm} that the
onset of deconfinement can be inferred 
through the dilepton 
yield if the quantity, $dN_{ee}/d\eta/(dN_{ch}/d\eta)^2$ 
(a quantity which is free from uncertainties in initial conditions)
reaches a plateau
with increasing $dN_{ch}/d\eta$ and the height of the plateau
will be a measure of $T_c$. We evaluated the above quantity 
for the four charge multiplicities mentioned above and did not observe 
any plateau.

\section{Summary and Discussions} 

We have studied the dilepton yield 
measured by CERES experiment
for various values of charge multiplicities
in Pb + Au interactions. It is 
observed that for lower multiplicities
($dN_{ch}/d\eta=150$ and $210$), dileptons seem to 
originate from a hadronic source with initial 
temperature $\sim 170$ MeV. 
For the higher values of the charge multiplicity (270 and 350), however,
the data can be described by both QGP and hadronic
initial states. The value of the initial
temperature realized in the high multiplicity
events is similar to that obtained from the analysis
of WA98 photon spectra. We find that both the enhancement
in the photon and dilepton yields measured by WA98 
and CERES collaborations respectively are due to a
thermal source with initial temperature $\sim 200$ MeV. 
However, it is difficult to state which one of the 
two phases, QGP or hadronic gas is realized in the
initial state of matter produced after the collision. 
We observe that the reduction of vector meson masses in the QHD model fails
to explain the low mass dilepton enhancement for $dN_{ch}/d\eta=350$.
We also point out that
the number of $\pi-\pi$ collisions extracted from the 
dilepton spectra near $m_\rho$  are reasonably high in order to
maintain thermal equilibrium in the system.  We conclude
by noting that in order to reproduce the data for large charge
multiplicities either a large broadening of $\rho$
or a substantial reduction in vector meson masses 
is required. Though it is difficult 
to draw a firm conclusion at this point,
these phenomena are closely related
to deconfinement or chiral symmetry restoring transition.
Indeed, in Ref.~\cite{kampfer} it has been shown that a parametrization 
which mimics the dilepton emission rate from $q\bar{q}$ annihilation
reproduce the CERES Pb+Au data well.
At this critical and interesting juncture we look forward to the 
experiments at RHIC where larger and hotter systems are likely to be produced.
Along with quantitative gains in all the signals there is a distinct
possibility of electromagnetic radiation from the QGP phase which will pave the way for an unambiguous conclusion regarding the formation of this novel
form of matter.

\noindent{\bf Acknowledgement:} We are grateful to B. Lenkeit
and J. Stachel for providing us with the experimental data. 
We also thank B. Sinha and T. K. Nayak for very fruitful discussions. 
J.A. is grateful to the Japan
Society for Promotion of Science (JSPS) for financial support.
J.A. and T.H. are also supported by Grant-in-aid for Scientific
Research No. 98360 of JSPS.

\end{document}